# Addressing Cognitive Biases in Augmented Business Decision Systems

Human performance metrics for generic AI-assisted decision making.

THOMAS BAUDEL*

IBM France Lab, Orsay, France

MANON VERBOCKHAVEN

IBM France Lab & ENSAE

VICTOIRE COUSERGUE

IBM France Lab, Université Paris-Dauphine & Mines ParisTech

GUILLAUME ROY

IBM France Lab & ENSAI

RIDA LAARACH

IBM France Lab, Telecom ParisTech & HEC

How do algorithmic decision aids introduced in business decision processes affect task performance? In a first experiment, we study effective collaboration. Faced with a decision, subjects alone have a success rate of 72%; Aided by a recommender that has a 75% success rate, their success rate reaches 76%. The human-system collaboration had thus a greater success rate than each taken alone. However, we noted a complacency/authority bias that degraded the quality of decisions by 5% when the recommender was wrong. This suggests that any lingering algorithmic bias may be amplified by decision aids. In a second experiment, we evaluated the effectiveness of 5 presentation variants in reducing complacency bias. We found that optional presentation increases subjects' resistance to wrong recommendations. We conclude by arguing that our metrics, in real usage scenarios, where decision aids are embedded as system-wide features in Business Process Management software, can lead to enhanced benefits.

---

* baudelth@fr.ibm.com.

# 1 INTRODUCTION

For the past 20 years, Business Process Management (BPM) [29] and related technologies such as Business Rules [10, 52] and Robotic Process Automation [36] have streamlined processes and operational decision-making in large enterprises, transforming work organization. These tools organize information circuits, from input events such as an order or a hiring request, through chains of stakeholders involved in various parts of the process, to reach some outcome. In a nutshell, a BPM system allows programming an enterprise like one would a robot. For some, these tools tend to rigidify the flow of information in a company and create a loss of context. However, there is ample evidence they provide numerous advantages: monitoring the flows of information, rearchitecting processes more easily, and providing much needed transparency and reliability to large and complex business entities. These technologies are used for instance in hiring processes, financial transactions oversight, procurement, business controls…

Because all inputs and decisions are stored, recent improvements involve applying machine learning techniques on past inputs and outcomes, to automate decision processes or to assist decision makers by providing suggestions on the likely outcome of each decision, assuming similar causes produce similar effects [39,51]. This technology ought to be largely beneficial, reinforcing the consistency of decisions and improving productivity, enabling "extended cognition", or "active externalism" as described by Chalmers [14]. Now, augmenting decision processes is not without risks. There is a large institutional community focusing on the area of AI Ethics, that stresses the requirements of fairness, transparency, explainabilty, accountability [28]… In particular, the prevention of algorithmic biases is a major concern [42], which needs to be addressed by technology [6], best practices [3] and regulations [37]. There is less institutional visibility on the changes to work practices and human decision-making these tools introduce. Cognitive biases induced by decision support systems are largely studied, identifying patterns such as automation and complacency biases [5] or, on the contrary, algorithm aversion [9] and decision fatigue [27, 4] when higher productivity is expected. There is less work directly applicable to the context of business decision-making, in particular in our present situation, where decision aids can be provided as a generic, system-level feature, regardless of their relevance for the task being performed by the human agent.

To ensure this type of assistance can be safely and profitably incorporated in business decision support systems, we first narrow down the type of tasks we are interested in augmenting, and the type of aids we wish to evaluate. Then, we review the literature to guide our designs. We present an empirical study of a simple decision task, which enables measuring the performance improvements augmented decision-making may provide, but also quantify various biases reducing the rationality of the subjects. We propose the definition of a *resistance* metric, that quantifies the ability of the human to retain their rationality. In a follow-up study, we assess the impact of 5 design choices to reduce those biases and increase resistance. Finally, we propose a methodology to incorporate bias measurement and compensation in business decision support systems, to ensure they provide their expected benefits with minimal inconvenience.

# 2 AUGMENTED BUSINESS DECISION MAKING

A business process is modeled in a flowchart (Figure 1). In this model, decision steps take as input some information, leverage external information, such as regulations or resource constraints, presumably only known to the decision maker, to advance the process through a predetermined set of possible outcomes. The type of decision tasks we investigate are constrained, leave little room for creative solutions (short of changing the process on the fly!) and presumably rely on a combination of explicit rules and heuristics.

When the decision logic can be formally expressed, Business Rules Management Systems (BRMS) [52] can automate the decisions, deterministically. When sufficiently robust heuristics are available, scoring methods or more sophisticated algorithms may also automate the decision, with an escalation process to handle exceptions. Finally, as in the hiring example, opening a new position in a specific team involves complex tradeoffs, which, we assume, involve so many factors that full automation is for now out of question. In these cases, a decision aid fed with previous decisions and possibly external data sources, such as revenue targets for the team and hiring requests in other teams, may provide information that can be useful to the general manager who decides to open a position.

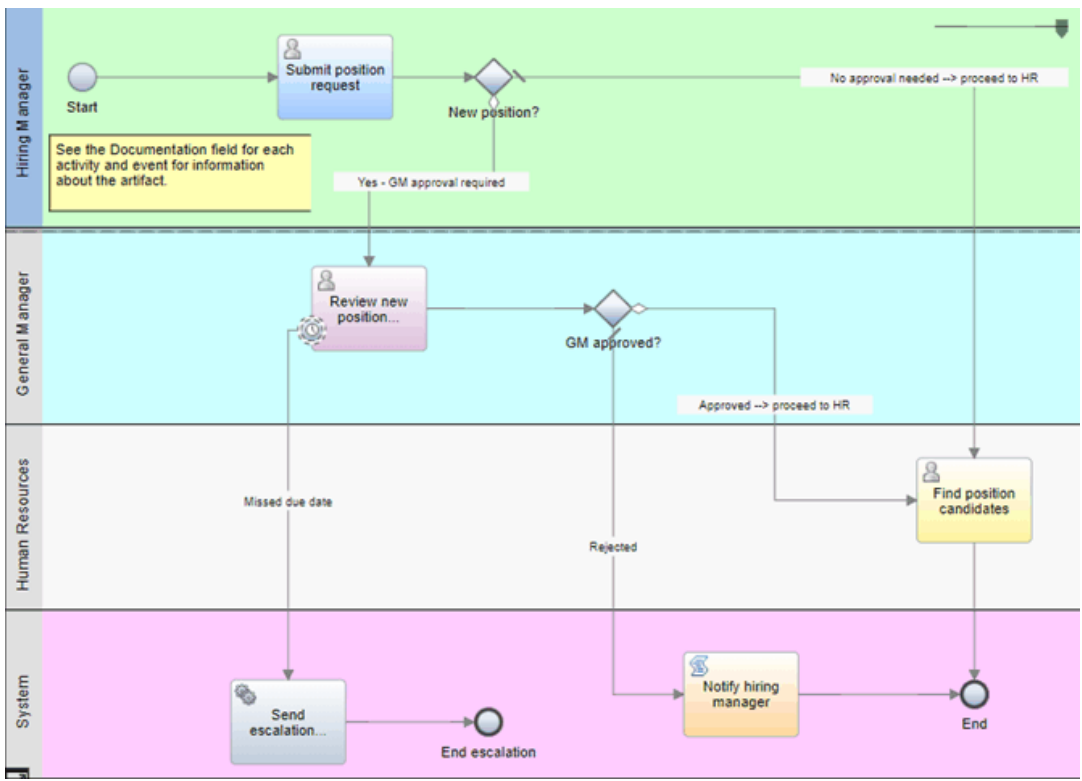

Figure 1: simplified hiring process represented as a workflow diagram. Horizontal bands represent the person or service involved in the process, rectangles depict tasks (which can be processes themselves), diamonds represent decision steps, which may be fully automated, or carried by humans, with or without decision aids.

More formally, we are interested in measuring, and possibly improve, the effects the presentation of decision aids may have in the following circumstances:
- Some information regarding a case is available and is assumed to be reliable.

- A choice, among a predefined set of alternatives, must be made to handle this case.
- The choice is partly constrained by some explicit constraints (regulations, resource limitations…)
- Some contextual information, exact or probabilistic, explicit or intuitive, is available to the decision maker (e.g. guidelines & priorities regarding the business context). Those allow the decision maker to form an *internal decision model*.
- Other information may be known only by the system, such as a history of similar cases.
- There is a best possible choice, but it may not be knowable in advance, if ever, for a given case.
- It is possible to provide, a posteriori, exact or approximate measures of which choices tend to perform better for which types of cases. These measures can be used to create a *computable decision model*.

There are numerous business situations that match this description, ranging from the mundane, such as a purchase decision, to more serious decisions such as hiring someone, granting a loan or selecting the winner of a bid, finally to morally heavy situations like deciding of a prison sentence [49]. Still, much decision-making activity, such as medical diagnostic or investment decisions, allow more creativity in choice-making and falls outside the scope of our work. What matters in our circumstances is that no algorithm may deliver a choice with 100% accuracy. An algorithm may be better than humans in general, but there is no substitute for human judgment and liability, when it is not possible, even a posteriori, to know if a a particular decision was the right one or not.

A variety of decision aids can be automatically provided in this context, which we categorize as follows, inspired from Miller's description of explanations [41]:
- Displaying more prominently attributes deemed most important in the decision (inputs)
- Scoring of the possible choices (computed by deterministic rules, deduction)
- Comparable cases and their outcomes (Nearest neighbors, induction)
- Decision tree classifier (such as random forests) output, its degree of confidence for the case, and the branch taken to provide the result (probabilistic deduction).
- Counterfactuals (if this attribute had this value, then the decision would be X with high certainty, abduction).

While there may be more types of decision aids, these cover the use cases we have reviewed in the literature that *can be implemented generically* (i.e. without specific tuning for the decision task, for instance creating a custom visualization for the decision step or for a given user). Finally, the presentation of these decision aids may influence their usage. They may be provided as plain recommendations, inciting the user to follow them, or available only on request, after a delay, or even after the decision has been made, as a verification step.

Our goal is to assess if decision aids improve the decision-making process, moving it towards a definition of rational decision-making suitable to our context. For now, we focus on performance metrics:
- Can we provide a methodology to measure decision-making performance in our contexts of use?
- Can the combination of human and algorithmic aids outperform both the human and the algorithm taken alone? How can we reach this stage of human-machine "collaboration"?
- Machine learning biases are a major concern. Even if this human-machine collaboration is an improvement, it seems inevitable that underlying algorithmic biases may taint decisions. Various cognitive biases may interfere. Can we identify and separate those to design correction strategies targeted at each of them?
- Can we define a measure of *resistance*, namely the ability for a human and an interface to reduce induced biases?

To address those questions, we carry an experimental study on a simple decision task meeting our requirements; we propose means to generalize this study in our contexts of interest, through the definition of some metrics. But before, we must acknowledge that there is a large body of literature addressing similar issues, which has guided our research.

## 3 RELATED WORK

### 3.1 Decision Theory

Understanding decision-making is a full research area in psychology. For a start, there are several positions regarding the notion of "correct" decision. For Rational Decision Theory, a rational choice maximizes an expected utility function [26, p.237]. Non rational theories [23] consider effects such as risk aversion or naturalistic viewpoints. Ultimately, these approaches can be reconciled when assumptions are clearly stated [30]. We take mostly a Rational Decision Theory standpoint, because we believe simple hypotheses are acceptable in our context. We also assume good will: the decision maker and the decision stakeholders (the company) share the same utility. Under stress, pressure, or poor motivation, we may find a divergence, which leads to a complacency bias or decision fatigue [4]. Within decision theory, our work falls into the area of judge-advisor systems [8]. Although much of the literature in this area is focused on human advisors, we retain the importance of advice presentation on the decision outcomes [38].

A major lesson of Decision Theory is that performance varies between individuals: experts and novices approach problems differently, and personality traits can have a strong influence [48]. Classical cognitive biases such as the order effect can be significant [47]. Finally, the availability of more information does not necessarily lead to better decision-making [17]. For fast and frugal or other recent approaches [25, 23], human decision making does not rely so much on *risk* - when the decision rules and probabilities of outcomes are well modeled - than on *uncertainty* - where the indecisiveness is not just a consequence of unknowns quantities, but also of unknows on the suitability of the decision process itself -. Hence, making a decision heuristically, based on limited information, may yield better results than paying close attention to possibly irrelevant details. Burton [9] convincingly defends that, by design, algorithmic decision aids operate under models of *risk*, while humans need to consider the *uncertainties* of a situation. Consciously leveraging this difference may provide the means to make the best of human and algorithm complementarity.

### 3.2 Decision Support Systems

Algorithmic decision aids have existed for a long time, in slightly different contexts.

*3.2.1 Semi-automation/process control*

Our early focus was on measuring, and possibly reducing, authority and complacency biases, which we felt should occur in our context. A large portion of the literature on these biases focuses on tasks that involve less dedicated attention and analysis than business decision. For instance, Parasuraman [44] finds attentional deficits at the onset of complacency biases. Bahner [5] identify this bias in process control tasks that involve verification rather than true decision making. Automation bias is clearly related to complacency bias. It can also result from attention deficits, which are even more prevalent in assisted driving tasks [21]. Alexander [2] finds a conformity bias: use by others increases trust in an algorithm. Finally, Gomboley [24], Yetgin [54] and Alberdi [1] describe a major cause of biases: when the algorithm outperforms the human most of the time, motivation necessarily dwindles. Conversely, Onnash [43] provides some evidence that decision aids lose all usefulness when they provide less than 70% accuracy, which indicates that providing those in generic business processes requires some prior assessment of relevance.

*3.2.2 Recommender Systems*

Recommender systems are algorithmic decision aids, but the tasks they support does not meet our focus: they don't help making a choice among predetermined outcomes, and decision quality is elusive: we often have to assimilate decision performance with user satisfaction [32]. Several types of cognitive biases can interfere, such as exposure bias [31]. Interestingly, recent literature in decision theory for recommender systems focuses on algorithm aversion, the opposite of complacency/authority biases: subjects avoid following recommendations, even when the algorithms perform better than humans [9]. The literature stresses the need to provide explanations [53, 46], especially for expert users [33]. We also notice that trust in the system degrades when bad recommendations occur [45], or when the task is highly subjective [13]. These findings should apply to our context of use.

*3.2.3 Visual Analytics for decision making*

Visual Analytics is an entire class of algorithmic decision aids. Visualization tools are geared towards exploratory analysis, which involves open decisions [20], thus is not in our focus. Still, work on identifying and reducing cognitive biases, such as the attraction effect [18] is relevant. More significant, the identification of the numerous biases [19] that may be found in visual decision aids provides a useful guiding framework.

**3.3 Impact of algorithmic decision aids on work practices, and ethical considerations**

*3.3.1 Medical decision support*

Medical decision support systems include decision aids and are extensively studied. Once again, they do not enter our scope, as the type of decision aids they provide are highly customized for specific purposes, and a medical decision can hardly qualify as a choice among predetermined possible outcomes. They support critical decisions, which explains why decision aids may be met with suspicion and some level of algorithm aversion [11]. Even in successful propositions [12], suspicion may not arise from the tool itself, but from the way it transforms work practices, perhaps for the better, but in directions which open avenues for high uncertainty [15]. At this point, addressing both algorithmic and cognitive biases in decision aids reaches beyond the scientific undertaking, into the realm of professional ethics. Determining the role of algorithmic decision aids requires a rigorous assessment of their power and their limitations in benefiting society.

*3.3.2 From scientific to ethical considerations*

Health professions are not the only industry questioning the impact of decision aids on work practices. Morris [50] devotes a whole chapter on cognitive biases in decision-making for accounting, and how overcoming those biases is an ethical issue. Legal and administrative professions raise similar concerns [16], including how proper explanations improve the perception of legal decisions [7]. Decision automation also produces surprising adaptive behaviors to circumvent the loss of control associated with algorithm-driven activities. For instance, [40] shows that human players learn to control computer players in computer games. Kyung Lee [35] describes how Uber drivers use collaboration to understand and regain some level of control over the dispatch algorithms. One of the possible shortcomings of algorithm-driven decision making (and decision aids) is the potential to induce unwanted behaviors by reverse engineering the decision aids' logic: the decision maker becomes 'controlled' by the subjects impacted by those decisions.

To address these concerns, expert groups and regulatory bodies provide guidelines [28, 3] to design trustworthy decision support systems. But in our context, we need more than design guidelines and regulations. We need metrics, and possibly a partially automated method to ensure those metrics stay within safe boundaries in the variety of contexts were algorithmic decision aids will be generically embedded in business decision support systems. As a first step towards this ambitious goal, we have conducted an empirical study on a simple task to guide the design of such metrics in the hope of generalizing them to real-world tasks.

## 4 STUDY DESIGN

Our study relies on a simple decision task: some information describing a case, a choice to be made among predefined possible outcomes, a right choice that depends on some rules and heuristics presumed known to the user (with an acceptable degree of variability). We test a single decision aid, presented as a recommendation from a generic classifier, under a variety of presentation modes. Sometimes the recommender will be misleading. We want to measure the success rate of the subjects, the impact of "wrong" recommendations – defining a notion of resistance - as well as other measures such as decision fatigue or time taken to reach a decision. The deviation from rationality is attributed to an authority or complacency bias, which are indistinguishable in our case. Both denote that the subject choses to follow the recommender over their own reasoning or intuition, in the first case because of a lack of motivation, in the second because they trust the algorithm more than themselves. Once we have obtained measures for a control condition (without decision aids), we provide decision aids with various presentations and observe the performance variations.

### 4.1 Choice of a decision task to evaluate

Resorting to a real-world task to define and assess our metrics is difficult: it requires extensive domain knowledge, and access to many expert users. Instead, we propose a simple decision task, for which a large population can create their own *internal decision model* easily. We leverage the well-known Titanic dataset[1], a database of passengers on the Titanic ship that sunk in 1912. This dataset is widely used to teach classification algorithms, because it exhibits some obvious patterns: most women in 1st and 2nd class survived, while most men in 3rd and 2nd class died, and other attributes such as number of relatives on board have a significant but lesser influence on the fate of each passenger. Simple machine learning classifiers, as well as humans after a short study of the data, exhibit success rates of 70-80% in guessing correctly if a passenger survived or not.

---

[1] https://www.kaggle.com/c/titanic

The decision task consists in deciding if a given passenger has survived or died. The task is repeated 20 times. To create an incentive, the presentation is somewhat gamified: subjects are enticed to maximize their score of correct guesses. Unbeknown to the subjects, the passengers presented all follow the expected distribution of survivors: a logistics regression classifier has >70% chance of correct classification on those passengers. Hence, we are asking the subject to make a rational choice - maximizing their probability of scoring high -, not a chance guess. This means reaching a perfect score is possible and even likely. Obtaining less than 50% (less than random chance) is a sure sign the subject is not properly committed to the task.

## 4.2 Stimulus presentation

The experiment starts with a few demographic questions: age range, level of studies and type of studies (humanities, business, engineering/science or other). Then the subject is presented with the goal, as well as some interactive visualizations (treemap) that let them create an *internal decision model*. We do not present explicit decision rules so as not to taint subjects (Figure 2). Next, we introduce the task, indicating that the recommender (in the experimental condition) has about 76% success rate of guessing correctly. Then, we present the stimuli (Figure 3). In accordance to the stated success rate of the recommender, 5 times in the run of 20 trials, the recommendation is wrong: it says "survived" when the subject has died or vice-versa.

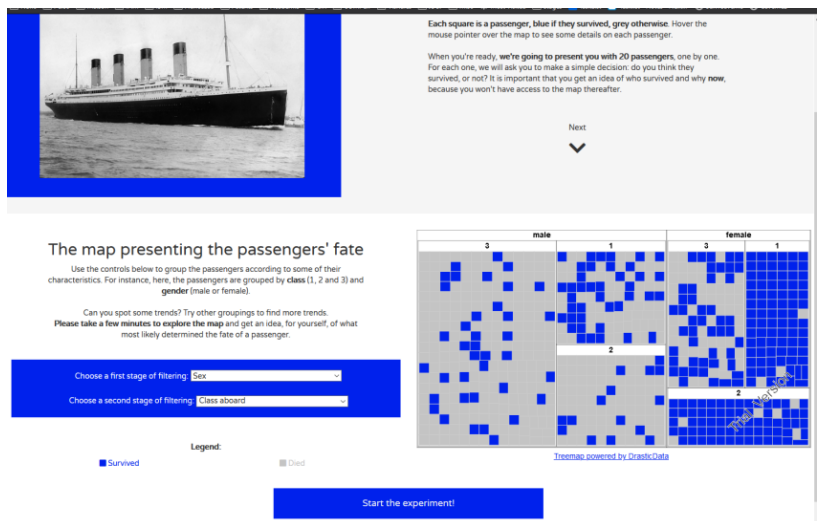

Figure 2: interactive visualization (treemap), helping the user form an internal decision model of who predominantly survived on the Titanic: Females from 1st and 2nd class, and who tended to die: males of 2nd and 3rd class.

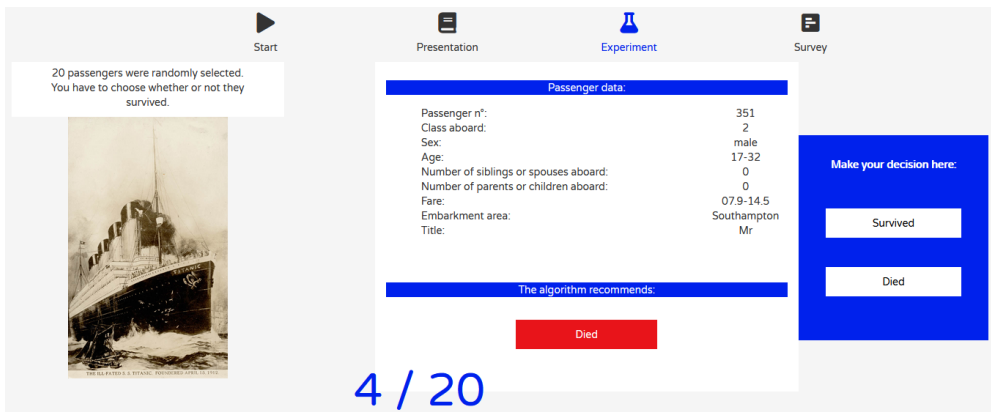

Figure 3: Stimuli in the experimental condition: passenger data, followed by a recommendation (dies or survives), and, on the right, 2 buttons "survives" and "die". In the control condition, the recommendation panel is not shown.

Finally, after 20 trials, we ask a few experience questions: estimated success rate, did they choose intuitively or using self-made rules (or don't know), how many times they think the recommender provided the wrong answer, and a free comment box. Finally, we provide their score, and an invitation to an event that presented the early results. We set a cookie on their browser so they can't repeat the experiment for a few hours, so as not to taint our data collection with repeated trials by the same subject.

### 4.3 Second run: presentation variants

The first run was meant to compare a control condition (no decision aid) with an experimental condition. In a second run, on a different population, we tested 5 different presentations:

- Control: the same presentation as in the first run
- Optional display: the recommendation is displayed only if the user requests it.
- Forced acknowledgement: the subject is presented with the passenger data, then must click a button to make the recommendation appear, and only then, after a small delay, they enter their choice.
- Reminder of 75%: instead of being only stated at the beginning, the subject is reminded that the decision aid has a 75% chances of success next to the recommendation.
- 80% success rate: instead of "guessing wrong" 75% of the time (5/20), the decision aid "fails" only 4 times out of 20. Of course, this is stated to the user, so it should not impact a rational behavior.

Each subject was assigned to the same condition for all their trials. The goal of this run was to assess how different presentation strategies and recommender reliability affect the measures of success, authority and resistance (defined below) in a significant way. This would be a strong indication that our metrics can be generalized to other context to drive the design of augmented business decision support systems.

### 4.4 Subjects recruitment and filtering

After a pre-run to calibrate expectations with ~20 subjects, we recruited subjects online. We did not want to use a survey service such as Amazon Mechanical Turk as we felt the subject's engagement would be distorted by a financial reward. Instead, we presented the experiment as a "fun and useful challenge". We announced the challenge on a variety of venues, starting in the company and student's forums and slowly extending our call to a wider audience, such as focused reddit and facebook groups, over a 2 months span.

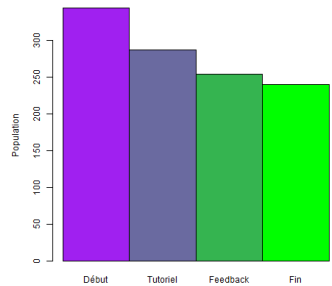

Figure 4: attrition for the second run: incoming participants, complete tutorial, reach feedback form, end page.

75% of the incoming participants completed the trials run, which we take as indicative of the motivation we had managed to induce from our anonymous subjects (Figure 4). A few participants (7) either failed to grasp the task or wanted to introduce noise and had less than 50% success (less than random choice), and we discarded them. In total, the first run had 231 participants and 155 usable trial runs, the second had 302 participants and 250 usable trial runs.

The demographics reached is a mix of students and educated professionals, with more of an engineering/science background, roughly equally distributed in the 20-55 age range (Figure 5).

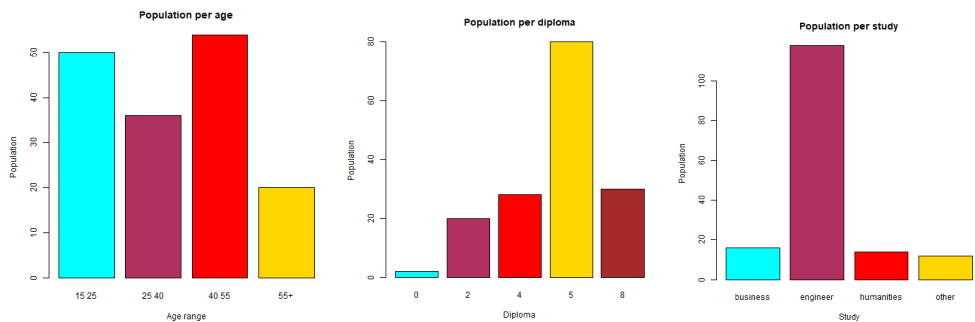

Figure 5: Distribution of the subjects by age range, years of study after high school, and type of study (self-reported).

### 4.5 Measures

To assess the impact of decision aid, we define a simple measurement of its effectiveness under a hypothesis of rationality: subjects want to maximize their success. Next, we define and measures of authority/complacency biases, i.e. how subjects deviate from rationality when they are subjected to misleading recommendations from the decision aid. For this, we perform three tests of increasing sophistication.

*4.5.1 Decision aid effectiveness & collaboration*

The most elementary measure is the effectiveness of the decision aids: does the combination of human and decision aid increase performance compared to each taken separately? This is quantified as the ratio:

$M_1$ = success in the experimental condition / max (success in the control condition, classifier success).

This ratio depends on a lot of factors:
- If a classifier clearly outperforms humans, there is little chance for this ratio to be > 1, full decision automation is most likely the best solution.
- Conversely, a classifier will likely be useless when its accuracy is < 70% [43]
- If the classifier and the humans tend to make the same type of mistakes, then there is no reason for $M_1$ to be above 1. In our setup, because the recommender choses to make a wrong recommendation at random, we may expect it to be above 1, unless some cognitive bias interferes.

This measure is easy to assess, explain, and it provides a nice metric to guide augmented Business Decision system design. It tells however little about how decision aids influence decision-making. The first, and perhaps most prominent bias we wanted to assess is authority/complacency bias: does a wrong recommendation influences the subject to let them lose their rationality? We have used 3 tests of increasing refinement to identify and measure this bias. These tests may be used to identify other cognitive biases when the experimental setup is relevant to exhibit them.

*4.5.2 Asserting authority/complacency bias with statistical modeling*

**Discrete modeling**: We can model the set of identical individuals and their independent responses as a Bernoulli variable. Considering the variable $Y \sim Bern(p)$ such that *P(Y=1) = P({decision is rational})*. We do not observe *Y*, but we know the distribution of the random variable *Y|{state_i}* where the states are:
- state 1: without recommendations; the control group.
- state 2: with recommendations: the treatment group.
- state 3: with correct recommendation (part of treatment).
- state 4: with incorrect recommendation (part of treatment).

We denote $Y_i$ for *Y|{state_i}*. Each variable $Y_i$ follows a distribution *Bern($p_i$)*. $Y_1$ and $Y_2$ are independent, but $Y_3$ and $Y_4$, coming from the same subjects, are not, thus they cannot be tested with the homogeneity test. We want to observe the effect of displaying a recommendation on the choice of the subject. If the subject's rationality is unaffected by the display of recommendations, we will have hypothesis H0: $p_1 = p_3$ and $p_1 = p_4$, else we will have H1: $p_1 \neq p_3$ and $p_1 \neq p_4$.

**Continuous modeling**: We can also consider a kernel density estimation. Given a random variable *P* representing the probability of a subject answering correctly. P(P<0.5) is the proportion of subjects with a success rate < 0.5. We consider $p_i$, a realization of P as an aggregate of responses (not individual responses).

Our trial runs provide a finite number of values for P. We use a sample drawn from a Gaussian kernel approximation of the test samples. We use the vector ($P_1$, $P_2$, $P_3$, $P_4$) for the estimation for P|{$state_i$}, while we have access to ($P_2$, $P_3$, $P_4$). We want to know if showing a recommendation will influence the rationality of subjects. If we note $F_i$ the cumulative density function of the rationality of group *I*, our hypotheses become: 1) Is the rationality impacted by the display of a good recommendation? H0: $F_1 = F_3$, H1: $F_1 \neq F_3$ and 2) is the rationality impacted by the display of a good recommendation? H0: $F_1 = F_4$, H1: $F_1 \neq F_4$.

Unlike the first model this allows us to compare the means of $P_3$ and $P_4$ with multivariate delta method. We will also test the following hypothesis: H0: $E[P_3] = E[P_4]$ and H1: $E[P_3] \neq E[P_4]$. To go from discrete values of P to continuous values, we use a Gaussian kernel estimation of bandwidth h = 0.1.

*4.5.3 Quantifying an authority/complacency bias with an econometric model*

Statistical modeling allows us to observe an effect, but not to quantify it. Econometric models provide us this ability.

**Defining and measuring authority/complacency bias**

In the following, subjects are indexed by *i*, while decision number are indexed by *t*. $Y_{i,t} = \mathbf{1}_{\{response\ of(I,t)\ is\ rational\}}$. We compare the control group with the treatment group that received wrong recommendations: $X_{i,t} = (1, \mathbf{1}_{\{false\ recommendaition\}})^T$. A linear panel model [22] is defined as:

$$Y_{i,t} = \mathbf{1}_{\{X_{i,t}^T \beta + \alpha_i + \epsilon_{i,t} > 0\}}$$

$$E[Y_{i,t}|X_{i,t}, \alpha_i] = X_{i,t}^T \beta + \alpha_i$$

$$\forall i \in \{1, ..., n\}, t \in \{1, ..., 20\}$$

$α_i$ corresponds to an individual effect. To estimate this model, we estimate:

$$Y_{i,t} = X_{i,t}^T \beta + \alpha_i + \eta_{i,t}$$

$$\mathbf{P}(\eta_{i,t} = 1 - X_{i,t}^T \beta | X_{i,t}, \alpha_i) = X_{i,t}^T \beta$$
$$\mathbf{P}(\eta_{i,t} = -X_{i,t}^T \beta | X_{i,t}, \alpha_i) = 1 - X_{i,t}^T \beta$$

Noting $v_{i,t} = α_i + η_{i,t}$, we have $\mathbf{E}[v_{i,s}\ v_{i,t}] \neq 0$, but $\mathbf{E}[v_{i,t}\ v_{i,t}] = 0$ because individuals are randomly assigned to the control and treatment groups and trials are randomly distributed. We can therefore estimate β with pooled OLS, which is a convergent estimator of β.

$$\beta_{MCO} = [\sum_{i=0}^{n}\sum_{t=0}^{t_i} X_{i,t} X_{i,t}^T]^{-1} \sum_{i=0}^{n}\sum_{t=0}^{t_i} X_{i,t} Y_{i,t}$$

The authority bias produced by a recommendation is the following quantity:

$$B(\alpha_i) = \mathrm{P}(\{Y_{i,t} = 0 | X_{i,t} = (1,1)^T, \alpha_i\} | \{Y_{i,t} = 1 | X_{i,t} = (1,0)^T, \alpha_i\})$$

In other words, it is the probability that a subject who sees a wrong recommendation will make a non-rational decision, considering he would have made a rational decision had he not seen this recommendation.

$$B(\alpha_i) = \frac{P(\{(1,1)\beta + \alpha_i \leq 0\}) \cap \{(1,0)\beta + \alpha_i > 0\})}{P((1,0)\beta + \alpha_i > 0)}$$
$$= \frac{(1,0)\beta + \alpha_i - (1,1)\beta - \alpha_i}{E[Y_{i,t}|X_{i,t} = (1,0), \alpha_i]}$$
$$= \frac{-\beta_2}{(1,0)\beta + \alpha_i}$$

**Defining and measuring resistance**

The resistance of a subject is his or her ability to overcome a wrong recommendation and stay rational, knowing he or she would have succeeded without a recommendation. Using the same econometric model, but on the subjects who received a wrong recommendation, we define $X_{i,t}$:

$$X_{i,t} = (1, \mathbf{1}_{\{wrong\_recommendation\}})^T$$

The ability of subjects to abstract themselves from the wrong recommendation is defined as:

$$C(\alpha_i) = P(\{Y_{i,t} = 1 | X_{i,t} = (1,1)^T\} | \{Y_{i,t} = 1 | X_{i,t} = (1,0)^T\})$$

It is the probability that the subject makes a rational decision when given a bad recommendation, knowing that if the recommendation had been correct, he would have made a rational choice.

$$C(\alpha_i) = P((1,1)\beta + \alpha_i + \epsilon_{i,t} > 0 | (1,0)\beta + \alpha_i + \epsilon_{i,t} > 0 | \alpha_i)$$
$$= \frac{P(\{(1,1)\beta + \alpha_i + \epsilon_{i,t} > 0 | \alpha_i\} \cap \{(1,0)\beta + \alpha_i + \epsilon_{i,t} > 0) | \alpha_i\}}{P((1,0)\beta + \alpha_i + \epsilon_{i,t} > 0 | \alpha_i)}$$
$$= \frac{P((1,1)\beta + \alpha_i + \epsilon_{i,t} > 0 | \alpha_i)}{E[Y_{i,t}|X_{i,t} = (1,0)^T, \alpha_i]}$$
$$= \frac{\beta_1 + \beta_2 + \alpha_i}{E[Y_{i,t}|X_{i,t} = (1,0)^T, \alpha_i]}$$

## 5 RESULTS

### 5.1.1 Decision aid effectiveness

The first run shows a significant collaboration effect: subjects in the control condition obtain a score of 72.3%, while subject in the experimental condition (with decision aid) have a 76% success rate, giving **M$_1$ = 1.014**.

|  | coefficient | 95% confidence interval |
| --- | --- | --- |
| Control condition (human alone) | 0.7230 | [0.6948, 0.7512] |
| With decision aid | 0.7604 | [0.7530, 0.7682] |
| "Algorithm alone" | 0.75 | -- |

Table 1: decision aid effectiveness (first run result)

While this improvement may appear modest, it should be reminded that our experiment is tuned to obtain average scores in the realistic range of 70%-80%. This collaboration effect is most useful to compare the effects of decision aids presentation synthetically. Our second run shows some modest but consistent variations:

|  | coefficient | $M_1$ |
|---|---|---|
| Control condition (human alone) | 0.7230 | 1 |
| With decision aid (new run) | 0.7651 | 1.020 |
| Optional display | 0.7655 | 1.020 |
| Forced acknowledgment | 0.7660 | 1.021 |
| Reminder of 75% | 0.7619 | 1.016 |

Table 2: measures of collaboration for various presentation modes of the recommender

Finally, in the condition where the recommender has a 80% success rate, we have a coefficient of 0.7822 (p-value < $10^{-4}$). Hence the collaboration effect disappears (**$M_1 = 0.977$**), indicating that automatic classification would be better suited to the task. While we have not tested a recommenda1tion with 70% or less success rate, we can assume from [43] that we would find a similar negative impact on the collaboration.

These measures of $M_1$ provide allow deciding which type of presentation and decision aid to chose for a given task and a given effectiveness of the algorithm. However, the cost of wrong decisions, particularly if they involve an algorithmic bias, may not be symmetrical: false positives and false negatives may have different costs in different scenarios. Hence, the measure of biases is important to assess the full cost/benefit analysis of choosing appropriate decision aids.

*5.1.2 Presence of an authority/complacency bias*

| Model | Test | H0 | H1 | P value |
|---|---|---|---|---|
| Bernoulli | $X^2$ homogeneity | $p_1 = p_3$ | $p_1 \neq p_3$ | 0.001 |
| Bernoulli | $X^2$ homogeneity | $p_1 = p_4$ | $p_1 \neq p_4$ | 0.047 |
| Bernoulli | Fit | $p_1 = p_3$ | $p_1 \neq p_3$ | 0.001 |
| Bernoulli | Fit | $p_1 = p_4$ | $p_1 \neq p_4$ | 0.047 |
| Bernoulli | $X^2$ independence | $Y_3, Y_4$ ind. | $Y_3, Y_4$ not ind. | < 0.001 |
| Gaussian kernel | Delta | $E[P_3] = E[P_4]$ | $E[P_3] \neq E[P_4]$ | < 0.001 |
| Gaussian kernel | Kolmogorov-Smirnov | $F_1 = F_3$ | $F_1 \neq F_3$ | < 0.001 |
| Gaussian kernel | Kolmogorov-Smirnov | $F_1 = F_4$ | $F_1 \neq F_4$ | 0.032 |

Table 3: results of the statistical tests.

All the tests reject the null hypothesis H0 to 5%, only hypothesis $p_1 = p_4$ and $F_1 = F_4$ require a threshold of 5%, for the others, 1% is sufficient. We conclude that the presence of a recommendation, and whether this recommendation is right or wrong, affects the rationality of the subjects, and the distribution functions of rationality are significantly different in the three cases.

- the probability of a rational decision is not the same depending on whether the user is alone, assisted by an incorrect recommendation or a correct one
- the distribution function of the degree of rationality is different for the three conditions.
- the success rate of subjects with an incorrect recommendation is not in-dependent of the success rate of subjects with a correct recommendation.

*5.1.3 Quantification of the authority/complacency bias (econometric model)*

Coefficients $β_1$ and $β_2$ of the two selected models are significantly nonzero at 5%. We can therefore reject the hypothesis that displaying a recommendation has no effect on the rationality of subjects. The model provides us with a metric that can be used to various groups of the pool of subjects and compare their relative rationality ($B(α_i)$, 0= little influence) and resistance ($C(α_i)$, 1=maximal resistance).

We can apply this model to study trends between different demographic classes recorded at the start of each run or presentation variants, between individuals, between trials (passengers) or any other available criteria. For instance, we display here the authority bias and resistance by type of studies (Table 4).

| Study type | Authority bias $B(α_i)$ | Resistance $C(α_i)$ | 95% conf. $B(α_i)$, | 95% conf. $C(α_i)$ |
|---|---|---|---|---|
| Engineering/science | 0.0666 | 0.8581 | [0.0626, 0.0705] | [0.8152, 0.9011] |
| Business | 0.0708 | 0.8423 | [0.0663, 0.0753] | [0.7898, 0.8947] |
| Humanities | 0.0684 | 0.8471 | [0.0641, 0.0726] | [0.7970, 0.8971] |
| other | 0.0737 | 0.8423 | [0.0687, 0.0787] | [0.7900, 0.8946] |

| Age range | $B(α_i)$ | $C(α_i)$ | Level of study | $B(α_i)$ | $C(α_i)$ |
|---|---|---|---|---|---|
| 15-25 | 0.0661 | 0.8606 | 2- | 0.0688 | 0.8553 |
| 25-40 | 0.0687 | 0.8473 | 4 | 0.0681 | 0.8552 |
| 40-55 | 0.0663 | 0.8522 | 5 | 0.0657 | 0.8537 |
| 55+ | 0.0713 | 0.8548 | 8 | 0.0694 | 0.8545 |

Table 4: authority bias and resistance for different demographic groups.

While the bias and resistance differences are small, those measures can be useful to apply to varying levels of expertise on a real task. Data and detailed results are available in the supplementary material.

*5.1.4 Comparison between presentation variants*

To choose the most effective presentation mode, depending if our goal is to maximize collaboration effectiveness, or to minimize authority bias while maintaining a high collaboration effectiveness, we compare the distributions of success under several conditions in Figure 6.

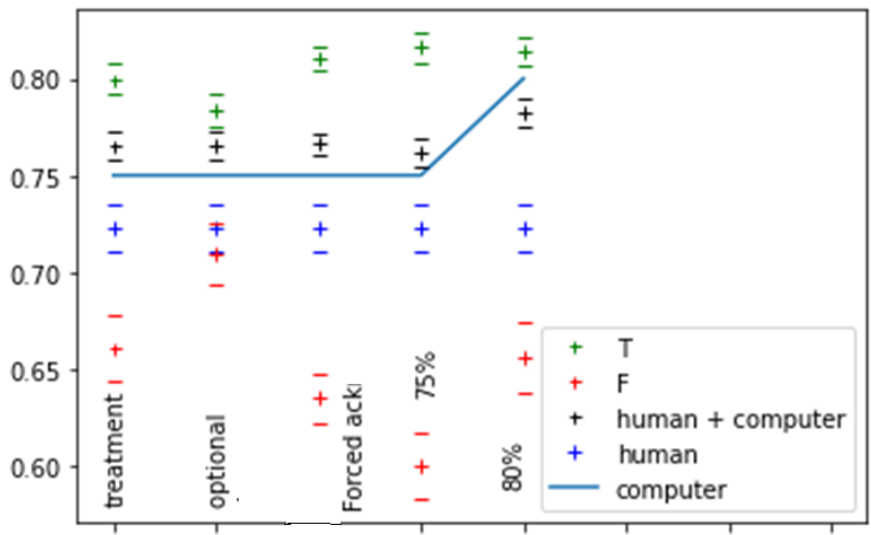

Figure 6: comparison of success rates with various stimuli presentations (black marks), success rates with wrong recommendation (red marks), success rate with good recommendations (green). Blue marks indicate the success without decision aid, and the line represent the "success rate" of the algorithm alone.

In Figure 6, we see success rate in the control condition as a horizontal line of blue marks, as a reference. Also for reference, the continuous horizontal line marks the "success rate" of the algorithm taken alone. The black marks indicate the success rate of the presentation mode. If the black marks are above both the blue marks (human alone) and the line ("computer alone"), then we can say the decision aid is effective in improving decision performance. This happens in all conditions but the last one, where the recommender has a much higher "success rate" of 80%. Based on this figure, the best performance is achieved with the "Forced acknowledgement" presentation mode (the subject must click to see the recommendation, then they can make their decision), although other presentations are quite close. This is the same information as tables 1 and 2.

But more importantly, the red marks show the distribution of success with a wrong recommendation. We can see that in the "optional display" presentation mode, the authority bias is weaker. This suggests using the presentation mode "optional display" over "Forced acknowledgement" if one is particularly concerned about avoiding underlying algorithmic biases.

Finally, our econometric model can provide similar measures for various subject groups, allowing to identify differences by type of participants (Figure 7). Here we see that the "75" presentation mode consistently induces more authority bias, while "Forced acknowledgment" induces less authority bias. The differences between groups appear not significant, except the 25-45 age group that seems to have slightly lower resistance.

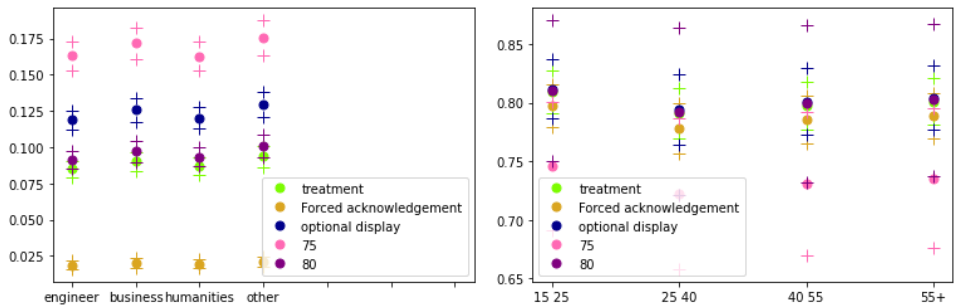

Figure 7: Authority bias by type of studies and presentation mode, and resistance bias by type of studies and presentation modes. These graphs show few differences between demographic groups.

*5.1.5 Qualitative feedback*

**Time spent** on this experiment is not relevant enough to justify elaborate statistics: we needed a task that could be completed fast to reach a large population. Aggregated data shows that, by and large, our assumptions hold. The average time to answer a trial is 10.3s in the control condition vs. 9.5s in the 1st experiment condition, with medians at 6.3s and 5.4s respectively, suggesting a small performance improvement with the decision aid. In the experiment condition, there is a slightly longer average time (9.6s vs 9.1s) when the recommendation is wrong than when it is correct, suggesting that subjects perceive the need to reflect when presented with a counter-intuitive proposition. Performance time by demographic category does not vary much, and finally, the time spent on the experiment influences very little the success rate (Figure 8).

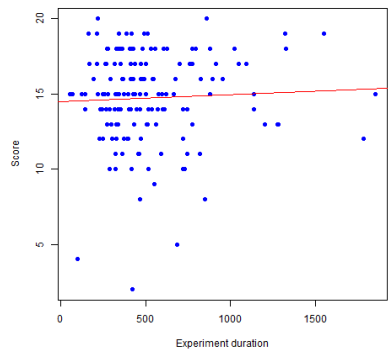

Figure 8: 2D plot time spent x success rate.

**Subjects feedback**

Subjects showed a variety of reactions. Some explained their reasoning: "i only looked at sex and class, taking more things into account was too confusing", "The pattern is pretty obvious. First class was a high chance of survival. Female is a high chance of survival. Children did better than adults. I didn't trust that the algorithm would do better than saying survive for 1st class and female.".

Others detailed their frustration at various constraints we had voluntarily set for the task: "I wish the algorithm had provided me with some explanation about its recommendations. Typically, when I disagreed with the recommendation, I would have loved to ask "why do you recommend this?"", "If the algorithm is only 80% accurate, why show us the algorithm answer before we make our decision?", "I would have needed a few explanations on how the algorithm works before the experiment and while doing it some feedback on how it decided would have been helpful.", "I would have preferred to answer with probabilities rather than a binary choice.", "Very confusing why i have to rate my own success rate but can't pick 50/50. Is the study about confidence? I don't recall a question asking my gender, so you may be unable to control for lower confidence bias in women...", "I would be interested to know the AI learning method".

Finally, many showed an understanding and appreciation for the study, noting that wrong recommendations could indeed affect their judgment: "Interesting experiment. Would love to see the end study!", "Funny (not topic, itself) and interesting", "AI = Random?", "I am very curious to understand the analyses process and the results. Would it be possible to receive the paper when published? Thanks". In debriefing interviews, subjects indicated that the test had made them aware of the complexity of the thought process at play when deciding to trust an algorithmic recommendation, which indeed, was the primary motivator of this study.

## 6 DISCUSSION

### 6.1 Limitations

Our measurements on this task may not be generalizable to other decision tasks. Our results align with [43], [24], [54] and [1]. Taken together, they suggest that decision aids are useful only when the "algorithm alone" success is within a certain range, which we can roughly estimate as[70%..human success rate + constant]. Still, the causes of the uncertainty in a decision-making task may vary widely, the bias patterns may equally vary. Our contribution lies in the metrics of collaboration, authority bias and resistance, how to assess them and put them in production for our specific context, it is not a contribution to Decision Theory.

The task we have evaluated is not an expert task, it is more comparable to a routine managerial decision than to a complex decision such as medical diagnosis. Even though the metrics we have defined can be applied to those more complex contexts, experimental setup and access to many experts should make this very difficult, justifying more longitudinal approaches such as [12]. As mentioned in the introduction, our focus is in providing generic decision aids in the context of business decisions, and our task matches this context.

Finally, the effects we have observed, while statistically significant, may seem quite modest. We have shown a significant advantage of augmented decision-making in certain circumstances, and a significant difference in several presentation modes to contain, or, on the contrary, increase, authority bias and resistance. These modest effects reflect the narrow window in which augmented decision-making has its usefulness. As our results suggest, when the algorithm clearly outperforms humans, it is probably better to rethink the role of the human. Conversely, when the algorithmic aid is of poor relevance, it is likely not useful, and, on the contrary, may lead to unwanted propagation of algorithmic biases through authority or complacency cognitive biases.

### 6.2 Performance metrics for augmented decision making

Our metrics enable the semi-automated introduction of performance measurement in decision support interfaces, which are appearing in many business processes. First, a measure of collaboration/decision aid effectiveness enables deciding if a decision support system provides value to the instrumented task, if the decision aid may just as well be removed to avoid introducing biases, or, on the contrary, if the task may be automated. Next, a measurement of authority bias and resistance based on an econometric model for different presentation modes enables refining these measurement to take into account asymmetries in the cost model of decision-making. It should be noted that other biases, such as decision fatigue, can be modeled using exactly the same tool, provided the evaluated task and data collection methods support this measurement. Initially, we wanted to measure this bias, but on a decisions that takes less than 10 seconds, any effect is difficult to exhibit.

Ultimately, a decision support tool performance should be modeled after an equation similar to:
$$V = (1-E) V_d - E C_d - C_t$$
Where E is the error rate of the decision task, $V_d$ is the value attributed to a correct decision, $C_d$ is the cost of making a wrong decision and $C_t$ is the cost of making the decision (typically human time and amortized time to develop the decision aid solution). The actual equation in a given situation should be weighted more precisely though, depending on various other costs, such as a difference between false positives and false negatives, or keeping decisions homogenous rather than maximizing value. Only a careful introduction of the metrics we propose can allow such fine-grained assessment of the benefit of decision aids in a particular context.

### 6.3 Towards a methodology and embedded measures of performance in augmented decision making

Our mid term goal is to apply the metrics we have defined to real usage scenarios, in several use cases we have identified with customers, such as the decision to start an audit on financial transactions that raise an alarm. Another important direction is to apply our methodology to other decision aids, such as nearest-neighbors methods, that shows the data and outcomes of cases closely related to the present case.
In the longer run, we envision that decision aids will be generalized in business decision systems, provided they are instrumented with tooling that continuously assesses their relevance and performance so as to mitigate risks while improving the productivity and quality of decision-making.

### 7 CONCLUSION

We have presented a set of metrics to include in A/B testing of decision aids used in business decision tasks to assess their usefulness and control the biases a particular decision aid and its presentation may induce on the decision maker. Applied to a simple decision task, our metrics can be used to show the possibility of human-system collaboration (72% of success for humans alone, vs 76% for a human assisted by an algorithm that gives the correct answer 75% of the time). We have also defined measurements of authority/complacency biases and defined a metric of resistance to this bias. Applied to our decision task, we found a significant effect of wrong recommendations on the rationality of subjects (-5%), indicating that underlying algorithmic biases in the decision aid may be propagated in the decision process instead of compensated by the human.

Testing several presentation variants, we found that a technique that presenting the recommendation only on request (optional) was effective in increasing the resistance of the subjects, all the while preserving the performance of decision-making.

Our measurement system is meant to be embedded in A/B testing of generic decision support system used in many contexts of business decision management systems and business processes. Augmented decision systems shift part of the responsibilities of the decision maker to the system designer. Introducing systems that may induce authority, complacency or other cognitive biases must be avoided. Hence, we believe furthering this work is important to shape the future of augmented business decision-making.

**ACKNOWLEDGMENTS**